\documentclass[12pt]{article}
\usepackage[dvips]{graphics}

\newcommand{\bd}[1]{ \mbox{\boldmath $#1$} }
\begin{document}
\def\ii{\'\i}

\title{A schematic model for QCD, II: finite temperature regime.}

\author{
S. Jesgarz \thanks{e-mail: jesgarz@nuclecu.unam.mx},
S. Lerma H.\thanks{e-mail: alerma@nuclecu.unam.mx},
P. O. Hess$^1$ \thanks{e-mail: hess@nuclecu.unam.mx}, \\
O. Civitarese$^2$ \thanks{e-mail: civitare@fisica.unlp.edu.ar},
and M. Reboiro$^2$
\thanks{e-mail: reboiro@fisica.unlp.edu.ar},
\\
\\ {\small\it$^{1}$Instituto de Ciencias Nucleares, Universidad
Nacional Aut\'onoma de M\'exico,} \\
{\small\it Apdo. Postal 70-543, M\'exico 04510 D.F.} \\
{\small\it$^{2}$ Departamento de F\'{\i}sica, Universidad Nacional de La Plata, } \\
{\small\it c.c. 67 1900, La Plata, Argentina. } \\
$\;\;\;\;\;\;\;\;\;\;$\\ }
\maketitle

\noindent {\small {\bf Abstract}: A schematic model for QCD,
developed in a previous paper, is applied to calculate meson
properties in the high temperature (up to 0.5 GeV) regime. It is a
Lipkin model for quark-antiquark pairs coupled to gluon pairs of
spin zero. The partition function is constructed with the obtained
meson spectrum and several thermodynamical observables are
calculated, like: the energy density, heat capacity, as well as
relative production rates of mesons and absolute production rates
for pions and kaons. The model predictions show a qualitative
agreement with data. Based on these results we advocate the use of
the model as a toy model for QCD.
} \\


\section{Introduction}

In Ref. \cite{paperI} (hereon referred to as (I)) a simple model,
representative of QCD, was introduced and applied to the
calculation of the spectrum of mesons. It is a Lipkin type model
\cite{lipkin} for the quark sector, coupled to a boson level which
is occupied by gluon pairs with spin zero. The four parameters of
the model were adjusted in order to reproduce 13 known meson
states with spin zero or one. The calculated spectra, for mesons
with spin different from the ones used in the fit, were found to
be in qualitative agreement with data. As reported in (I), the
calculated meson states contain many quarks, antiquarks and
gluons. The gluon contributions were found to be of the order of
30$\%$. The model predictions (I) are free of the so-called
multiplicity problem, i.e. that a given state can be described in
many ways, which is removed due to the action of particle mixing
interaction. The model itself resembles the one of Ref.
\cite{schutte} which treats  nucleons coupled to pions. Also, it
is related to the work of Ref. \cite{pittel}, which describes
quarks and uses particle conserving interactions. Generally
speaking, the model of (I) belongs to the class of models
described in Refs. \cite{stuart,betabeta}. The gluon part in (I)
is fixed \cite{gluons99} and does not contain any new parameters.
The validity of the basic theoretical assumptions, and the
applications to low and high temperature regimes, has been studied
for mesons with flavor (0,0) and spin 0 \cite{simple}. The aim of
these studies was to formulate a manageable, schematic, albeit
realistic, model to describe qualitatively QCD at low and high
energies. Since the model is algebraic, i.e. all matrix elements
are analytic, and exactly solvable, it can provide a
non-perturbative description based on QCD relevant degrees of
freedom, like quarks, antiquarks, and gluons. This, in turn,
allows to test other microscopic many body techniques previously
applied to the non-perturbative treatment of real QCD
\cite{swan,bonn}. Although the proposed model (I) is probably too
simple to describe real QCD, it contains all basic ingredients of
real QCD. These are the correct number of degrees of freedom
associated to color, flavor and spin, and the orbital degree of
freedom , which is contained in the degeneracy $2 \Omega$ of each
of the quark levels.

In this work we investigate the behavior of the model, in the
finite temperature regime. By starting from the model predictions
of the meson spectrum, we calculate the partition function and
different thermodynamical quantities, like the energy density and
the heat capacity as a function of temperature. Next, we focus on
the calculation of meson production rates. As we shall show, these
production rates are in qualitative agreement with the
experiments. Also, we calculate absolute production rates for
pions and kaons. Finally, we concentrate on the transition from
the Quark-Gluon-Plasma (QGP) \cite{qgp,qmd} to the hadron gas. The
results support the notion that the present model may be taken as
a toy model for QCD.

The paper is organized as follows: In section 2 the model is
shortly outlined, since the details have been presented in (I). In
section 3 we calculate the partition function and give the
expressions for the relevant observables. In section 4 the model
is applied to the description of the QGP. There, we present and
discuss the results corresponding to some branching ratios and
absolute production rates. Finally, conclusions are drawn in
section 5.

\section{The model}

As described in (I), the fermion (quarks and antiquarks) sector of
the model consists of two levels at energies  $+\omega_f$ and
$-\omega_f$, each level with degeneracy $2\Omega$ $=$ $n_cn_fn_s$,
where $n_c=3$, $n_f=3$ and $n_s=2$ are the color, flavor and spin
degrees of freedom, respectively (see Fig. 1 of (I)). Each level
can be occupied by quarks. Antiquarks are described by holes in
the lower level. Equivalently, one can use only the positive
energy level and fill it with quarks and antiquarks with positive
energy. The Dirac picture is useful because it gives the
connection to the Lipkin model as used in nuclear physics. The
quarks and antiquarks are coupled to gluon pairs with spin zero.
The energy of the gluon level is 1.6 GeV \cite{gluons99}, and the
energy $\omega_f$ is fixed at the value $\omega_f=0.33$ GeV ,
which is the effective mass of the constituent quarks.

The basic dynamical constituent blocks of the model are
quark-antiquark pairs $\bd{B}^\dagger_{\lambda f, SM}$  which are
obtained by the coupling of a quark and an antiquark to flavor
$\lambda $ ($\lambda = 0,1$) and spin $S$ ($S=0,1$). The index $f$
is a short hand notation for hypercharge $Y$, isospin $I$ and its
third component $I_z$. Under complex conjugation the operator
obeys the phase rule defined in \cite{draayer}.

The states of the Hilbert space can be classified according to the
group chain

\begin{eqnarray}
[1^N] &  [h]=[h_1h_2h_3] & [h^{\rm{t}}] \nonumber \\
U(4\Omega )  & \supset  U(\frac{\Omega}{3})~~~~~~~~~~ \otimes & U(12) \nonumber \\
&  ~~~ \cup ~~~ & \cup \nonumber \\
&  (\lambda_C,\mu_C)~SU_C(3) ~~~~  (\lambda_f,\mu_f) & SU_f(3) \otimes SU_S(2) ~S,M ~~~,
\label{group1}
\end{eqnarray}
where the irreducible representation (irrep) of the different
unitary groups are attached to the symbols of the groups. The
irrep of $U(4\Omega)$ is completely antisymmetric (fermions) and
the one of $U(\frac{\Omega}{3})$, the color group $U(3)$ for
$\Omega =9$, and $U(12)$ are complementary \cite{hamermesh}. The
color irrep $(\lambda_C,\mu_C)$ of the color group $SU_C(3)$ is
related to the $h_k$ via $\lambda_C=h_1-h_2$ and $\mu_C=h_2-h_3$.
The reduction of the $U(12)$ group to the flavor ($SU_f(3)$) and
spin group ($SU_S(2)$) is obtained by using the procedure
described in \cite{ramon,sergio}. In (\ref{group1}) no
multiplicity labels are indicated (see (I)).

The classification appearing in (\ref{group1}) is useful to
determine the dimension and content of the Hilbert space. Instead
of working in the fermion space we have introduced a boson mapping
\cite{klein,hecht}. The quark-antiquark boson operators are mapped
to

\begin{eqnarray}
\bd{B}_{\lambda f, S M}^{\dagger} & \rightarrow &
\bd{b}_{\lambda f, S M}^{\dagger} \nonumber \\
\bd{B}_{\lambda f, S M} & \rightarrow &
\bd{b}_{\lambda f, S M}
~~~.
\label{bosons}
\end{eqnarray}
where the operators on the right hand side satisfy exact boson
commutation relations.

The model Hamiltonian is defined completely in the boson space and
it is given by

\begin{eqnarray}
\bd{H} & = & 2\omega_f \bd{n_f} + \omega_b \bf{n_b} + \nonumber \\
& & \sum_{\lambda S} V_{\lambda S}
\left\{ \left[ (\bd{b}_{\lambda S}^\dagger )^2 +
2\bd{b}_{\lambda S}^\dagger \bd{b}_{\lambda S} + (\bd{b}_{\lambda S})^2 \right]
(1-\frac{\bd{n_f}}{2\Omega})\bd{b} + \right.
\nonumber \\
& & \left. \bd{b}^\dagger (1-\frac{\bd{n_f}}{2\Omega})
 \left[ (\bd{b}_{\lambda S}^\dagger )^2 +
2\bd{b}_{\lambda S}^\dagger \bd{b}_{\lambda S}+ (\bd{b}_{\lambda
S})^2 \right] \right\} ~~~, \label{hamiltonian}
\end{eqnarray}
where $(\bd{b}_{\lambda S}^\dagger )^2$ $=$ $(\bd{b}_{\lambda
S}^\dagger \cdot \bd{b}_{\lambda S}^\dagger )$ is a short hand
notation for the scalar product. Similarly for $(\bd{b}_{\lambda
S})^2$ and $(\bd{b}_{\lambda S}^\dagger \bd{b}_{\lambda S})$. The
factors $(1-\frac{\bd{n_f}}{2\Omega})$ simulate the terms which
would appear in the exact boson mapping of the quark-antiquark
pairs. The $\bd{b}^\dagger$ and $\bd{b}$ are  boson creation and
annihilation operators of the gluon pairs with spin $S=0$ and
color $\lambda=0$. The interaction describes scattering and vacuum
fluctuation terms of fermion and gluon pairs. The strength
$V_{\lambda S}$ is the same for each allowed value of $\lambda$
and $S$, due to symmetry reasons, as shown in  (I). The matrix
elements of the Hamilton operator are calculated in a seniority
basis. The interaction does not contain terms which distinguish
between states of different hypercharge and isospin. It does not
contain flavor mixing terms, either. The procedure used to adjust
the four parameters (values of $V_{\lambda S}$), was discussed in
detail in (I).

The disadvantage posed by working in the boson space is the
appearance of un-physical states. In (I) we have presented a
method which is very efficient to eliminate spurious states, as we
shall show in this paper. The suitability of the Hamiltonian
(\ref{hamiltonian}) to describe the gluon pair and quark-antiquark
pair contents of mesons has been discussed in details in (I).

As a next step, in this paper, we have introduced temperature and
discussed the transition to and from the QGP. As it can be
expected, because of the schematic nature of the model, we may
attempt to describe only the general trends of the observables. To
achieve this goal, further assumptions have to be made respect to
the volume of the system because the model, as it has been
proposed in (I), has no a priory information about the volume of
the particle.

\section{The partition function, some state variables and observables}

The group classification of the basis (\ref{group1}), allows for a
complete book-keeping of all possible states belonging to the
Hilbert space of (\ref{hamiltonian}). The corresponding partition
function, which contains the contribution of the quark-antiquark
and gluon pairs configurations introduced in the previous section,
is given by

\begin{eqnarray}
Z_{qg^2_0} & = & \sum_{[h]}dim(\lambda_C,\mu_C)
\sum_{(\lambda_f,\mu_f)}
\sum_{J} (2J+1) \sum_{P=\pm} \sum_i \nonumber \\
& & mult(E_i) e^{-\beta (E_i - \mu_B B - \mu_s s - \mu_I I_z)}
~~~, \label{partition}
\end{eqnarray}
where $\mu_B$, $\mu_s$ and $\mu_I$ are the baryon, strange and
isospin chemical potentials, respectively. The sum over
$[h]=[h_1h_2h_3]$ denotes all color irreps of $U(3)$ with $\sum_k
h_k =N$, where $N$ is the total number of quarks in the two levels
(Dirac's picture). The transposed Young diagram $[h]^{\rm{t}}$,
obtained by interchanging rows and columns, denotes the $U(12)$
irrep. The index "i" refers to all states with the same color,
flavor, spin and parity ($P$). These states are obtained after the
diagonalization of the Hamiltonian (\ref{hamiltonian}). For mesons
belonging to the $\pi$-$\eta$ and $\omega$-$\rho$ octet, the mass
values entering in (\ref{partition}) do not take into account
flavor mixing. The eigenvalues $E_i$ are denoted by the eigenvalue
index $i$, and they are functions of all the numbers needed to
specify the allowed configurations, namely: $s$,
$(\lambda_f,\mu_f)$, $P$, $S$, $[h]$ and of the cutoff for the
different boson species $[\lambda , S]$ (see (I)). The quantities
$B$ and $I_z$ are the baryon number and the third component of the
isospin. According to the experimental evidences the value $\mu_I
=0$ is a reasonable approximation, and we have consistently
adopted it in our calculations. The dimension corresponding to
color configurations is given by $mult(\lambda_f,\mu_f)$ $=$
$\frac{1}{2}(\lambda_f+1)(\mu_f+1)(\lambda_f+\mu_f+2)$ and of the
spin by $(2J+1)$.

Since the eigenstates of the Hamiltonian (\ref{hamiltonian}) have
been calculated after performing a boson mapping, as described in
(I), we have consistently fixed the corresponding cutoff-values at
the values $2\Omega$, $\Omega$, $\frac{2\Omega}{3}$, and
$\frac{\Omega}{3}$ for the boson pair species [0,0], [0,1], [1,0]
and [1,1] respectively (see (I)). These values are adequate when
the fermion (quark-antiquark) configurations entering in the boson
states correspond to a full occupation of the fermion lower state
($-\omega_f$). These numbers may be modified when the fermion
configurations correspond to states where the upper level is
partially occupied and the lower level is partially unoccupied.
The distribution of the occupation in the upper and lower fermion
levels is fixed for a lowest weight state $|lw>$ of a given
$U(12)$ irrep, defined by $B_{\lambda f, S M} |lw>=0$. The irrep
of $U(12)$ is given by a Young diagram \cite{hamermesh} with $m_k$
boxes in the $k$'th row. The lowest weight state is given by
$\sum_{k=1}^6 m_k$ quarks in the lower level and $\sum_{k=7}^{12}
m_k$ quarks in the upper level. The highest weight state is
obtained by interchanging the occupation. The difference of the
number of quarks in the upper level, appearing in the highest and
lowest weight states, gives the maximal number of quarks we can
excite for a given U(12) irrep. This number is given by

\begin{equation}
2J = \sum_{k=1}^6 m_k - \sum_{k=7}^{12} m_k ~~~.
\label{j}
\end{equation}
For the case $[3^60^6]$, used in (I), we found $J = \Omega$.
Therefore, $2J$ is the maximal number of quarks we can shift to
the higher level, i.e. it is equal to the maximal number of
quark-antiquark pairs which can be put on top of the lowest weight
state of a given $U(12)$ irrep, which is also the state with the
lowest energy in absence of interactions.

The total partition function is given by

\begin{equation}
Z = Z_{qg^2_0} Z_{g} ~~~,
\label{ztot}
\end{equation}
where $Z_{g}$ is the contribution of all gluon states
\cite{gluons99} which do not include contributions of gluon pairs
with spin zero. It is written $Z_g= \sum_\alpha exp(-\beta
E_\alpha )$. The values $E_\alpha$ can be deduced by using Eq.
(40) of Ref. \cite{gluons99}. Except for the gluon pairs with spin
zero, all other gluon states are treated as spectators because the
Hamiltonian in (\ref{hamiltonian}) does not contain interactions
with these other states. Note, that the interaction between the
gluons is taken into account explicitly in the model of Ref.
\cite{gluons99}. As a short hand notation we will abbreviate the
partition function by $Z=\sum_i e^{-\beta {\cal E}_i}$, taking
into account that ${\cal E}_i$ contains the information about the
chemical potential, the contributions of the quarks and gluons.

The observables $<\bd{O}>$ are calculated via \cite{greiner}
\begin{eqnarray}
<\bd{O}> = \frac{ \sum_i \bd{O} e^{-\beta {\cal E}_i} }{Z_a} ~~~,
\label{average}
\end{eqnarray}
where the index $a$ denotes the color configurations, i.e;
$a$=(0,0) when only color zero states are considered and $a=c$
when also states with definite color are allowed. This distinction
is needed to investigate the phase where color confinement is
effective and the phase where color is allowed over a wide area of
space. The quantities to calculate are the internal energy
($<E>$), heat capacity ($<C>$), average baryon number ($<B>$),
strangeness ($<s>$) and the expectation value of different
particle species ($<n_k>$), where $k$ refers to the quantum
numbers of a particular particle and $\sum_{k} <n_{k}>=1$, with
the sum over all possible quantum numbers.

The particle expectation values have a simple expression because
they select one of the eigenvalues at the time, thus if the state
of a given particle is denoted by "$i$" and $E_i$ is its energy,
the particle expectation value is given by

\begin{eqnarray}
< n_i > & = & \frac{e^{-\beta (E_i - \mu_B B -\mu_s s)} }{Z_a} ~~~,
\label{n-exp}
\end{eqnarray}
(where we have used the value $\mu_I=0$, for the isospin chemical
potential ).

At this point we have to make an assumption upon the volume
considered. The whole reaction volume can be divided in elementary
volumes, and we assume that the elementary volume ($V_{el}$) is of
the size of a hadron, corresponding to a sphere with a radius of
the order of 1 fm. Later on we shall show that this choice is
reasonable, as seen from the calculated thermodynamic properties
of the whole system. Another assumption is related to the
interaction, which does not take into account confinement. We
shall discuss two scenarios, namely: a) no additional interaction
related to color is taken into account for temperatures above a
critical (de-confinement) value, and b) confinement is operative
for temperatures below the critical de-confinement temperature. In
the regime (a) the lowest state with color (1,0) lies at the
energy $2\omega_f$, and it corresponds to put one quark in the
upper fermion level. Although it is a possible configuration, the
Hamiltonian (\ref{hamiltonian}) cannot act upon it. This is
consistent with the fact that above a certain temperature, $T_c$
(de-confinement temperature), only color non-singlet states are
allowed. The actual value of $T_c$ will then give us an idea about
the regime where hadronization is operative. In the real world
hadronization, i.e. confinement, should set in below a critical
temperature, as a true phase transition. In our model this will be
signaled by a sharp transition from a state where color
non-singlet states are still allowed ($T>T_c$) and a state where
confinement is effective ($T\leq T_c$).

\section{Description of the high temperature regime: the QGP}.

We shall first discuss the case where no additional color
interaction is taken into account. The states with energy $E_i$
for a given flavor, spin and parity are obtained from the
diagonalization of the model Hamiltonian, now including flavor
mixing and the corrections due to the Gel'man-Okubo mass formula
for the two lowest meson nonets (one with spin zero and the other
with spin 1). In Figure 1 we show the internal energy as a
function of the temperature $T$, with and without interactions.
The results shown by a dashed line have been obtained by
calculating the internal energy in the fermion space {\it without}
interactions. In this case the Hamiltonian has a simple image in
the fermion space and the calculation of the partition function
can be performed exactly. The dotted line shows the results
obtained by working with the boson mapping, and by enforcing the
corresponding cut-offs in the maximal number of bosons as
explained above and in (I). The internal energy is a good
indicator of the number of active states in the Hilbert space.
Note that the results shown by both curves, the doted and dashed
lines, practically coincide. This suggests that the number of
active states is nearly the same in the boson and fermion spaces
for a wide range of temperatures. This does not imply that all
un-physical states have disappeared but rather that the
approximate method of cutting un-physical states works reasonable
well. The curve shown by a solid line, in the same Figure 1, gives
the internal energy obtained from the calculation performed in the
boson space and in presence of interactions. Although the curve
does not show a clear phase transition of first order (e.g: a
sharp increase of the energy in a narrow interval around $T_c$ )
the behavior around $T_c=0.170$ GeV is pretty suggestive of it. We
have interpreted the observed smearing-out of the curve as
follows: for $T=0$ the vacuum state is dominated by pairs of the
type [1,0] and in this channel a quantum phase transition
\cite{simple} does indeed take place. By this we mean that the
pairs [1,0] are effectively blocked at high temperature. The other
channels contribute less significantly to the ground state (see
(I)) and interact weakly than the ([1,0]) channels, therefore,
they remain in a perturbative regime. Thus, as the temperature
increases, an approximate first order phase transition takes place
in the channel [1,0], a mechanism similar to the one shown in Ref.
\cite{simple} for the [0,0] channel, while the other
configurations remain un-affected. The superposition of these two
mechanism leads to the smearing out of the curve around $T_c$, as
shown in Figure 1.

\begin{figure}[tph]

\rotatebox{270}{\resizebox{356pt}{356pt}{\includegraphics[100,130][612,660]{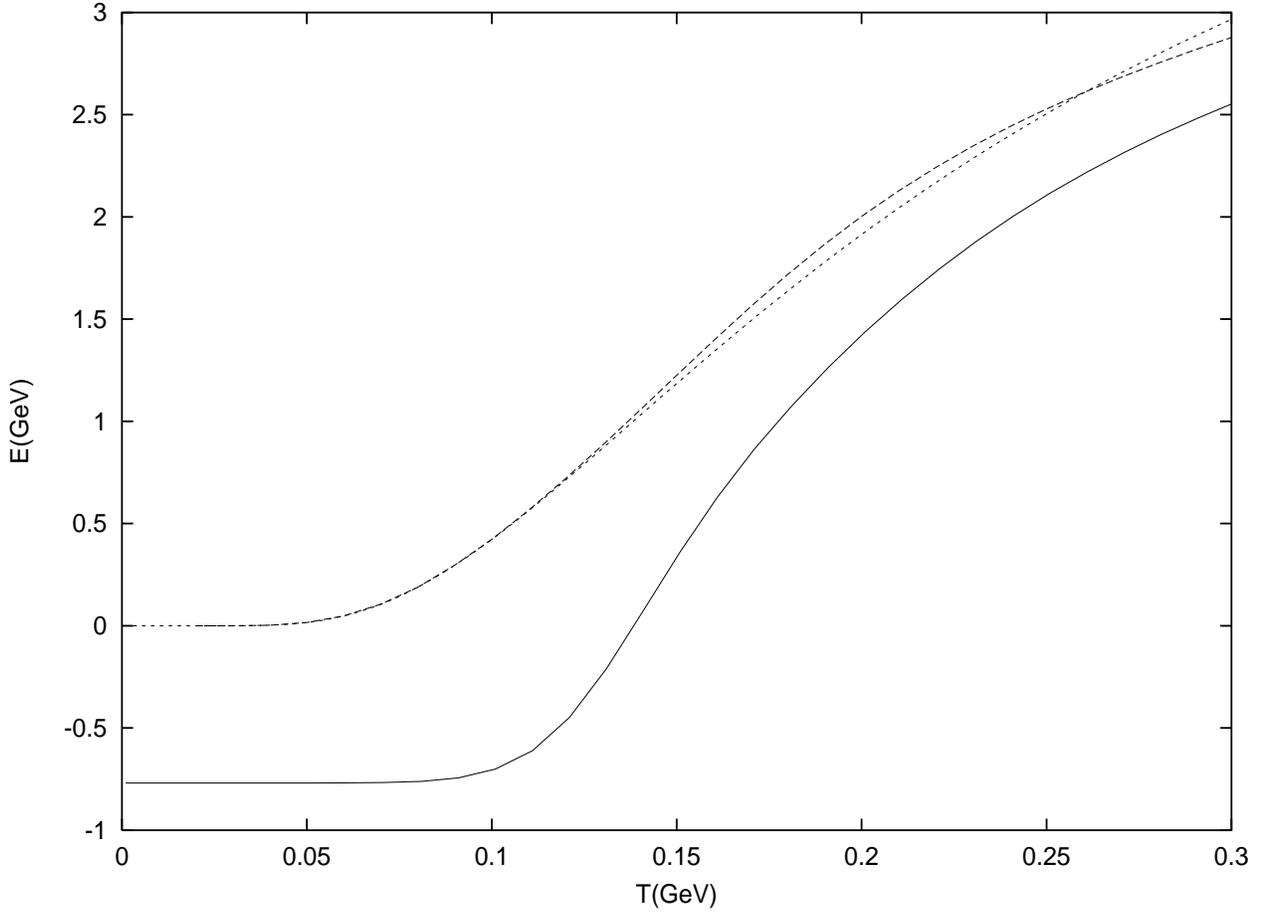}}}
\vskip 0.5cm \caption{ Internal energy as a function of $T$. The
dashed line corresponds to the exact result in the fermion space,
without interaction. The dotted line is the internal energy
calculated in the boson space with the cut-off for the different
boson species, as discussed in the text. The solid line represents
the results obtained with the full Hamiltonian and calculated in
the boson space. No additional color interaction is taken into
account. } \label{figure1}
\end{figure}

\begin{figure}[tph]

\rotatebox{0}{\resizebox{656pt}{656pt}{\includegraphics[200,130][612,660]{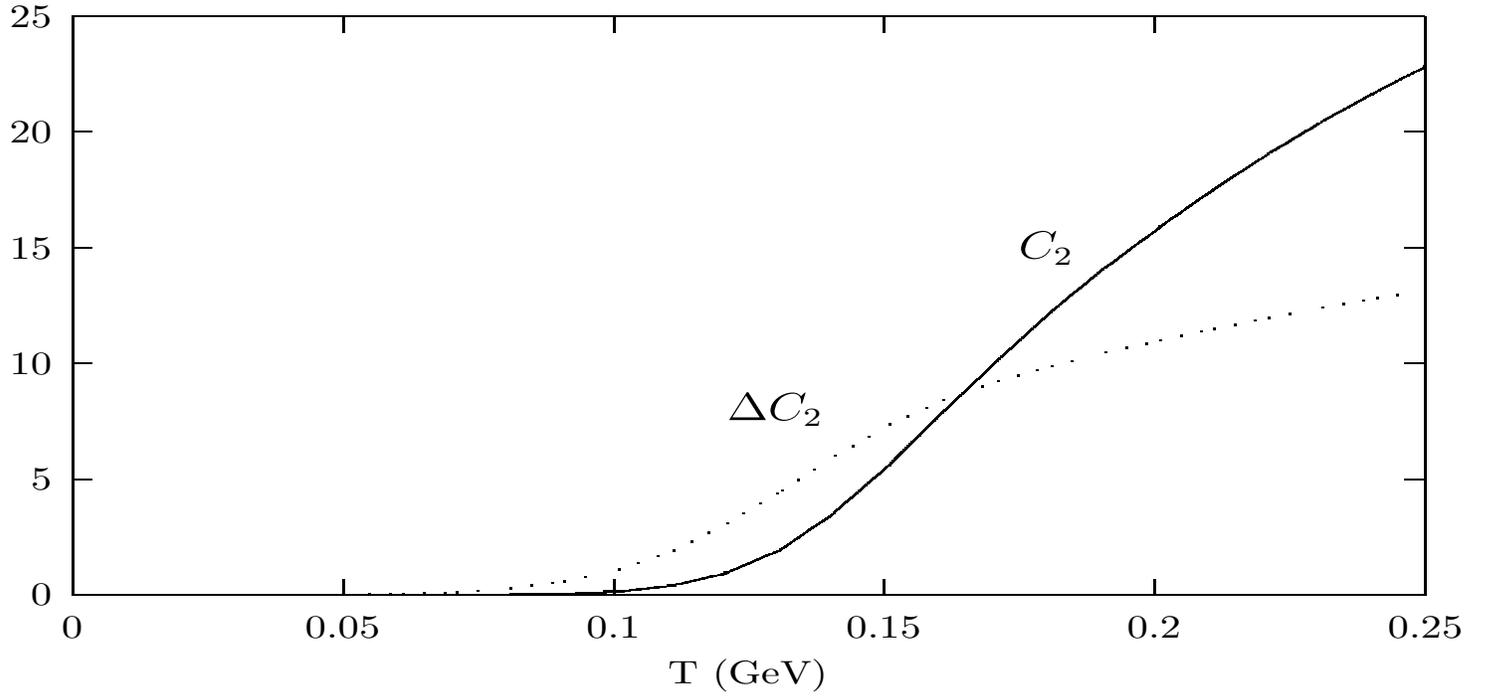}}}
\vspace*{-10cm} \caption{ The expectation value of color
($C_{2}=<\bd{C_2}>$, solid line) and its variation ($\Delta C_{2}
= \sqrt{<\bd{C_2}^2>-<\bd{C_2}>^2}$, dotted line) as a function of
$T$. The variation reaches the same value as the expectation value
around $T=0.170$ GeV. The values $\mu_B=\mu_s=0$ were used in the
calculations. } \label{figure2}
\end{figure}

In Figure 2 the expectation value of the Casimir operator
($C_2=<\bd{C_2}>$) of color and its variation ($\Delta C_2 =
\sqrt{<\bd{C_2}^2>-<\bd{C_2}>^2}$) are shown. The eigenvalue of
the Casimir operator, for an irrep with color numbers
$(\lambda_C,\mu_C)$, is given by $C_2(\lambda_C,\mu_C)=
\lambda_C^2$ $+$ $\lambda_C\mu_C$ $+$ $\mu_C^2$ $+$ $3(\lambda_C +
\mu_C)$. As a reference, for a color (1,0) irrep $C_2=3$ while the
irrep (1,1) has $C_2=9$. We assume that $\mu_B$ and $\mu_s$ are
zero. It is interesting to observe that, in the present model, the
variation of the color is approximately symmetric around $T=0.170$
GeV. A possible interpretation is the following: at high energy
the probability to have a color non-singlet state is large (the
variation is not large enough to allow color singlet states) and a
QGP is formed where color is effective over a wide range in space.
From $T=0.170$ GeV on the probability to find a state in color
(0,0) is significantly increased, since the variation is large
enough to allow color singlet states. In lowering the temperature
the variation is much larger than the average color and the whole
QGP dissolves in droplets of color zero. Within the present model,
these results, of the average color and its variation, are signals
of the transition to the hadronic phase. Accordingly, we assume
that it takes place near $T_c=0.170$ GeV, for $\mu_B=\mu_s=0$. We
may now calculate the bag pressure and construct the $T-\mu_B$
diagram. At $T=0.170$ GeV the pressure is determined via the
expression  $p=\frac{Tln(Z)}{V_{el}}$, where $\Phi = -Tln(Z)$ is
the grand canonical partition function \cite{greiner} and $V_{el}$
is the elementary volume $V_{el}=\frac{4\pi}{3}r_{el}$. For
$r_{el}$=1 fm we obtain a bag pressure $p^{\frac{1}{4}}$ of about
0.17 GeV, which is in reasonable agreement with standard values.
When the chemical potentials $\mu_B$ and $\mu_s$ are different
from zero, the temperature dependence of the internal energy
changes and also changes the value of the temperature for which
the pressure is equal to the bag pressure $p^{\frac{1}{4}}$ $=$
$0.18$ GeV. The results are shown in Figure 3. Assuming that the
local strangeness is $<s>=0$, we arrived at a functional relation
between $T$, $\mu_B$ and $\mu_s$, i.e. $f(T,\mu_B ,\mu_s )=0$,
which fixes $\mu_s$ as a function of $\mu_B$. The results of this
functional relation are displayed in Figure 4.

\begin{figure}[tph]

\rotatebox{270}{\resizebox{356pt}{356pt}{\includegraphics[100,130][612,660]{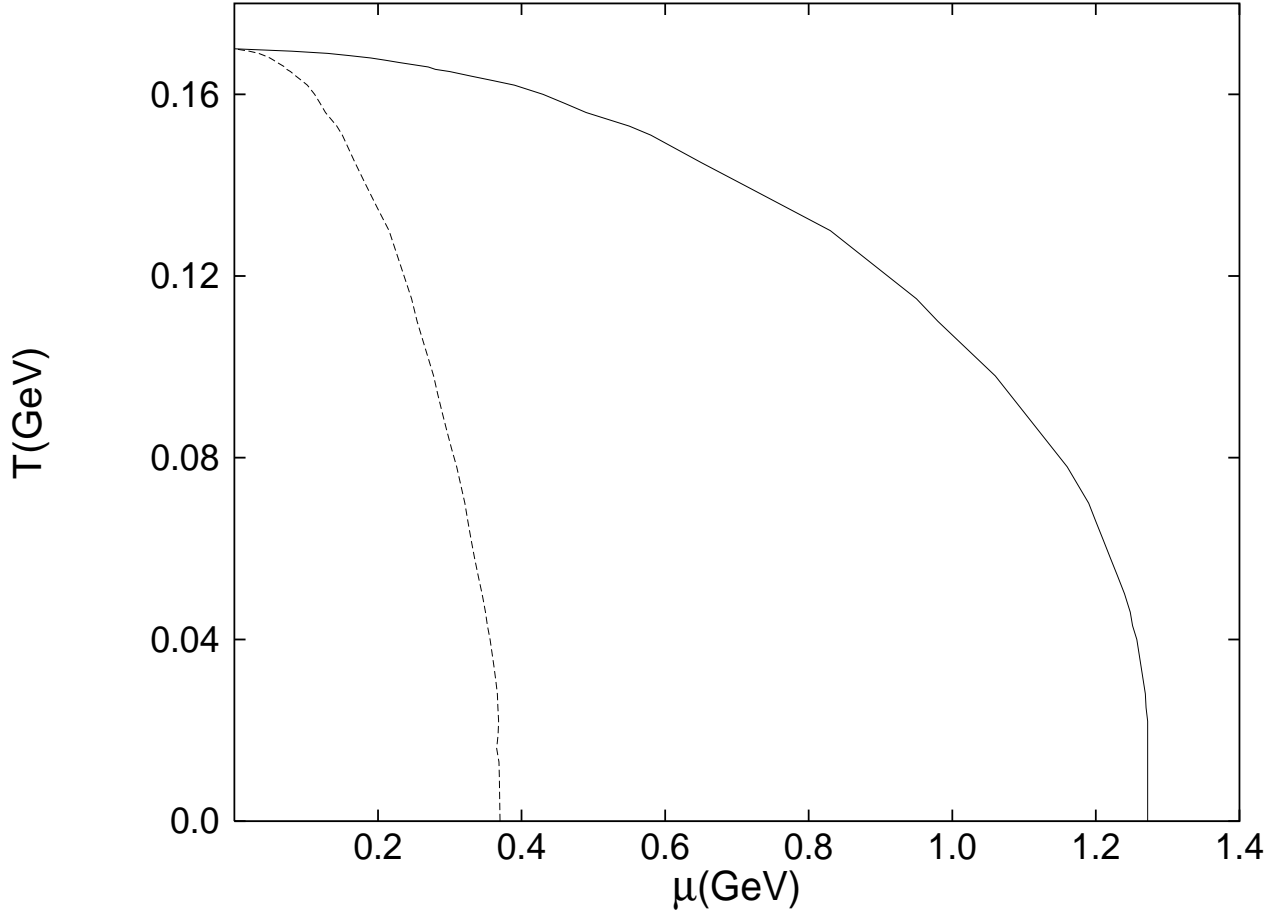}}}
\vskip 1.5cm \caption{ Temperature versus chemical potential. The
curves gives the boundary where the pressure of the system in an
elementary volume of radius $r=1$ fm is equal to the bag pressure.
The values of the temperature $T$ are given as a function of the
chemical potentials $\mu_B$ (solid line) and $\mu_s$ (dashed
line). The label $\mu$ denotes both $\mu_B$ and $\mu_s$. }
\label{figure3}
\end{figure}

\begin{figure}[tph]

\rotatebox{270}{\resizebox{356pt}{356pt}{\includegraphics[100,130][612,660]{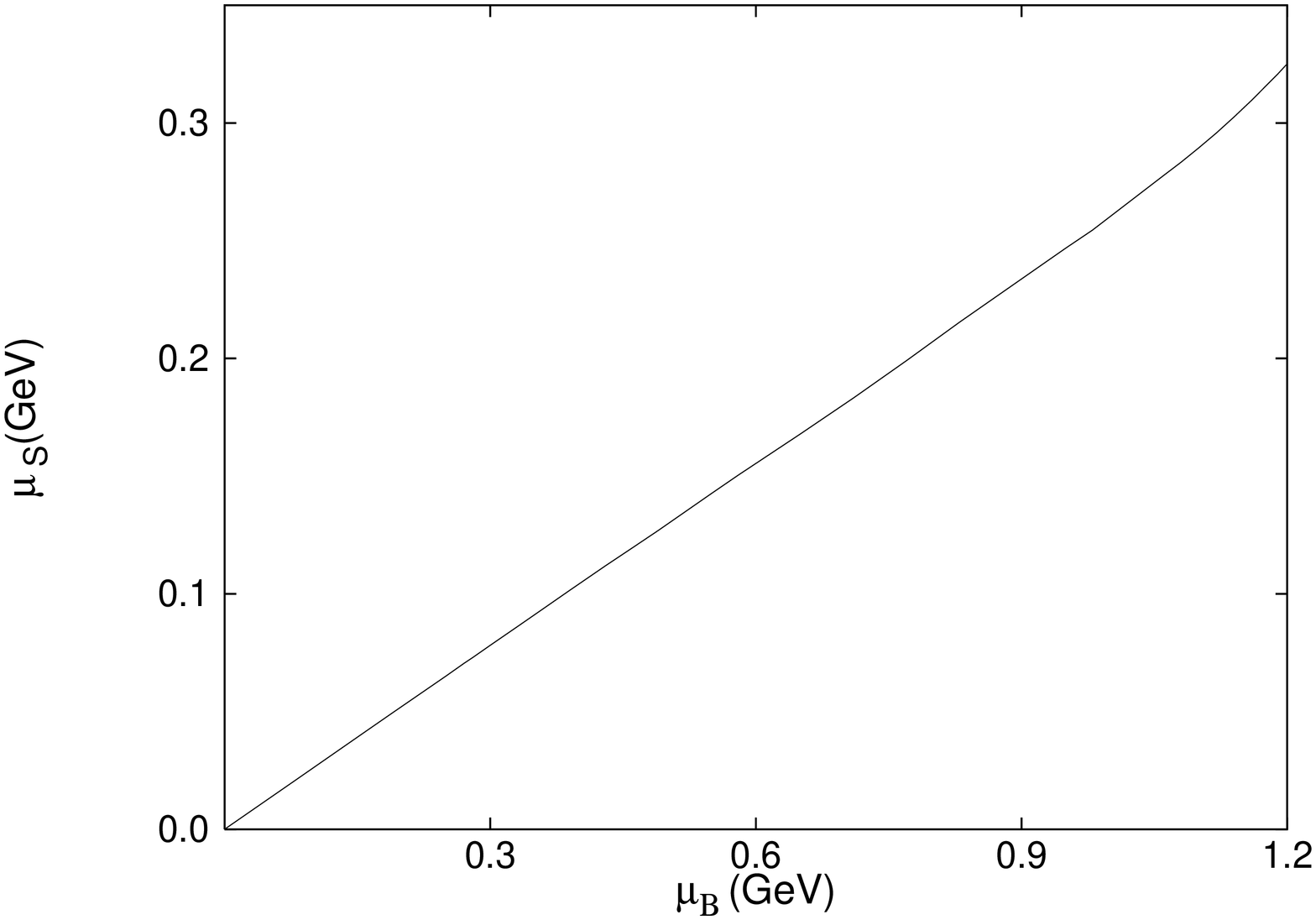}}}
\vskip 0.5cm \caption{ Chemical potential $\mu_s$ as a function of
the chemical potential $\mu_B$, for zero local strangeness
$<s>$=0. To each pair of values ($\mu_B$ and $\mu_s$) it
corresponds a temperature $T$, which is determined from the
limiting values shown in Figure 3.} \label{figure4}
\end{figure}
Once the chemical potential $\mu_s$ is adjusted, by using the
results shown in Figures 3 and 4, the chemical potential $\mu_B$
and the transition temperature $T_c$ can be consistently
determined.

Up to now we did not take into account an interaction which
generates confinement. This has to be done by hand. One
possibility is to {\it assume} that the transition from the QGP to
the hadronic phase takes place within a very small range of
temperatures around the critical temperature $T_c$. We require
that the partition function above $T_c$ allows any color while for
$T<T_c$ it contains only color zero states. Finally, chemical
equilibrium connecting both phases, the QGP and the hadron gas, is
understood. Figure 5 shows the results of the internal energy,
without confinement (upper curve) and with confinement (lower
curve). The solid line connecting both curves indicates the values
for which confinement vanishes above $T_c$. As seen from the
results, the transition is now of first order. Figure 6 shows the
heat capacity calculated for the case without confinement (dotted
line) and with confinement below $T_c$ (dashed line). The solid
line interpolates between them, as in the case of the internal
energy (see Figure 5).

\begin{figure}[tph]

\rotatebox{270}{\resizebox{356pt}{356pt}{\includegraphics[100,130][612,660]{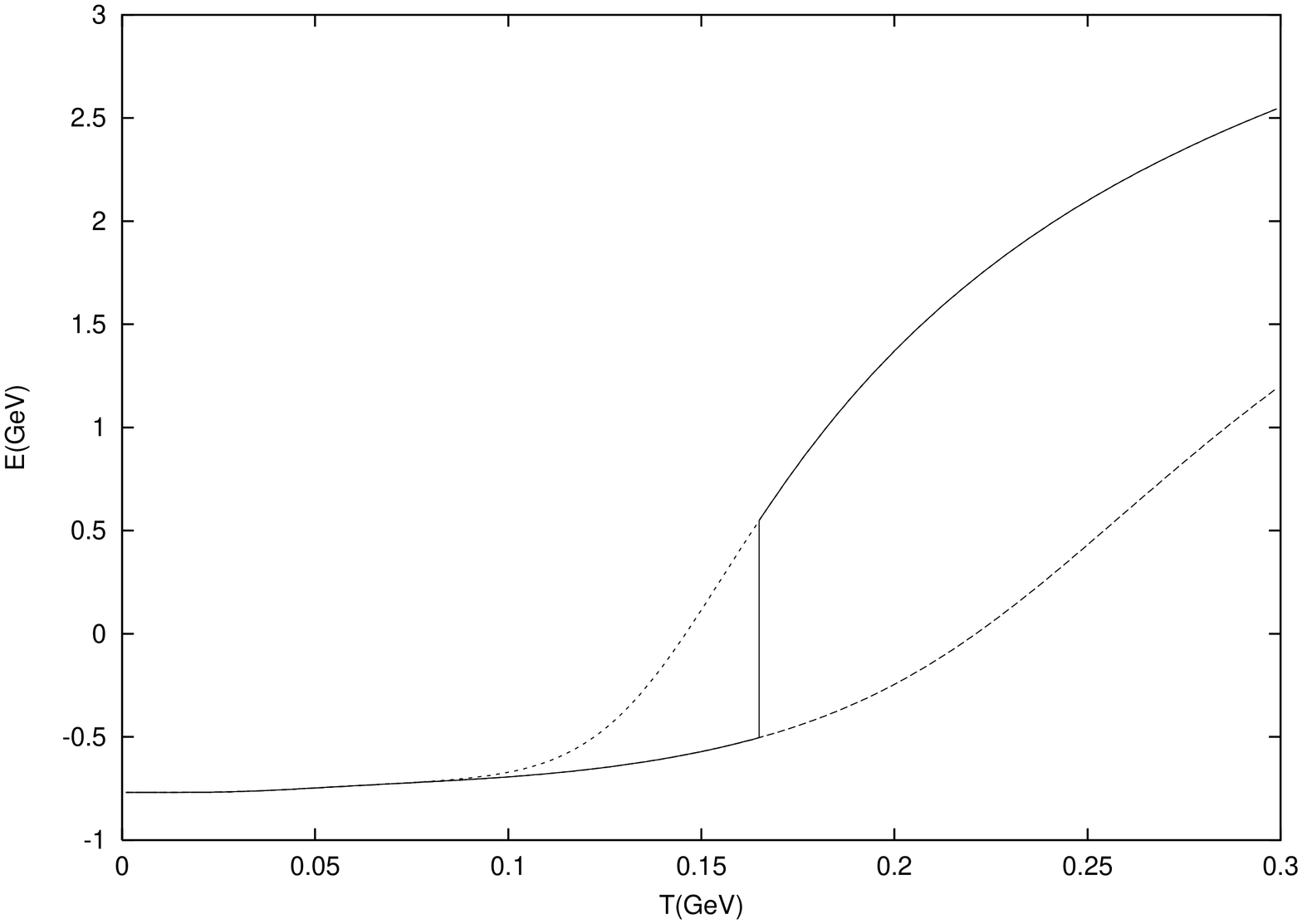}}}
\vskip 0.5cm \caption{ Internal energy. The upper curve shows the
results obtained without considering confinement. The lower curve
corresponds to the results obtained with the inclusion of
confinement. The solid line shows the results obtained with the
partition function where confinement is considered to be fully
operative below $T_c$ ($\mu_B$=$\mu_s$=0). } \label{figure5}
\end{figure}

\begin{figure}[tph]

\rotatebox{270}{\resizebox{356pt}{356pt}{\includegraphics[100,130][612,660]{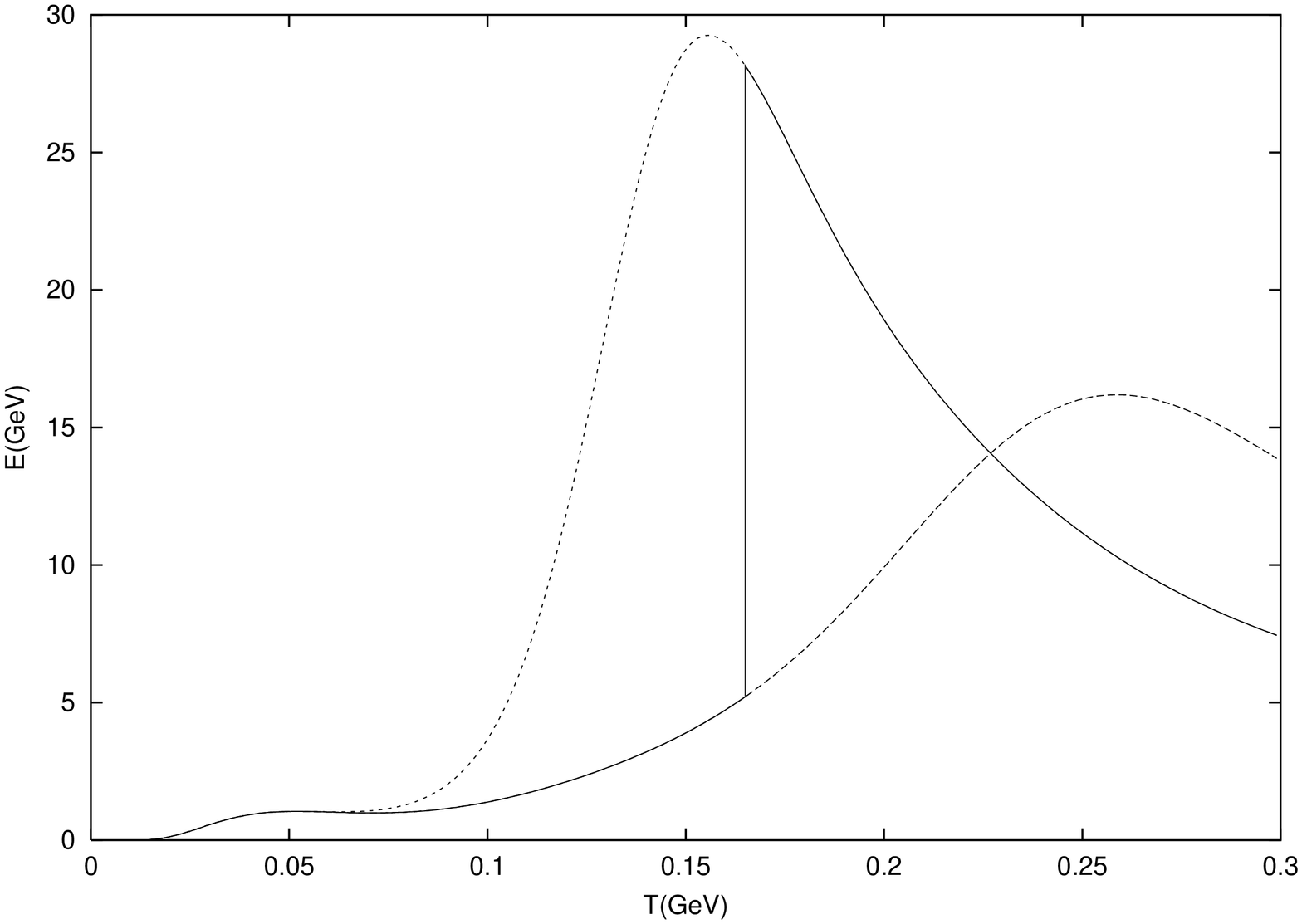}}}
\vskip 0.5cm \caption{ Heat capacity of the system. The solid line
interpolates between the case with confinement below $T_c$ to the
case without it above $T_c$ ($\mu_B$=$\mu_s$=0). The dotted and
dashed line correspond respectively to the cases where color is
allowed and without color. } \label{figure6}
\end{figure}

The model can first be tested in the energy region below the
transition temperature $T_c$, where the hadron gas should prevail.
The confinement is effective and therefore we have to use the
partition function $Z_{a=(0,0)}$ in the equations (\ref{average})
and (\ref{n-exp}). We take, as an example, the measured total
production rates of $\pi^+$ at 10 GeV/A, as reported in the
SIS-GSI experiment \cite{gsi-10gev}. The system considered was
Au+Au and we assume that all particle participate, i.e.
$N_{part}$=394. In Figure 2.3 of Ref. \cite{gsi-10gev} a
temperature of about $T=0.13$ GeV is reported. Assuming local
strangeness conservation , $<s>=0$, we obtain a relation of
$\mu_s$ versus $T$, which is depicted in Figure 7. From there we
obtain, for the reported temperature, $\mu_s \approx 0.128$ GeV
and via Figure 4 a value $\mu_B \approx 0.55$ GeV. In Figure 8 we
show for a fixed value of $\mu_B$, physically acceptable at
temperatures near $T=0.13$ GeV, the resulting total production
rate of $\pi^+$ ($N_\pi$) as a function in the temperature $T$.
For $T$ = 0.13 GeV the production rate is approximately 180 pions
$\pi^+$, which is close to the value 160, which we have obtained
by using Figure 2.3 of Ref. \cite{gsi-10gev}. The good qualitative
agreement with the experiment demonstrates that the present model
is able, indeed, to describe, approximately, observed QCD
features.

\begin{figure}[tph]

\vskip 1cm
\rotatebox{270}{\resizebox{356pt}{356pt}{\includegraphics[100,130][612,660]{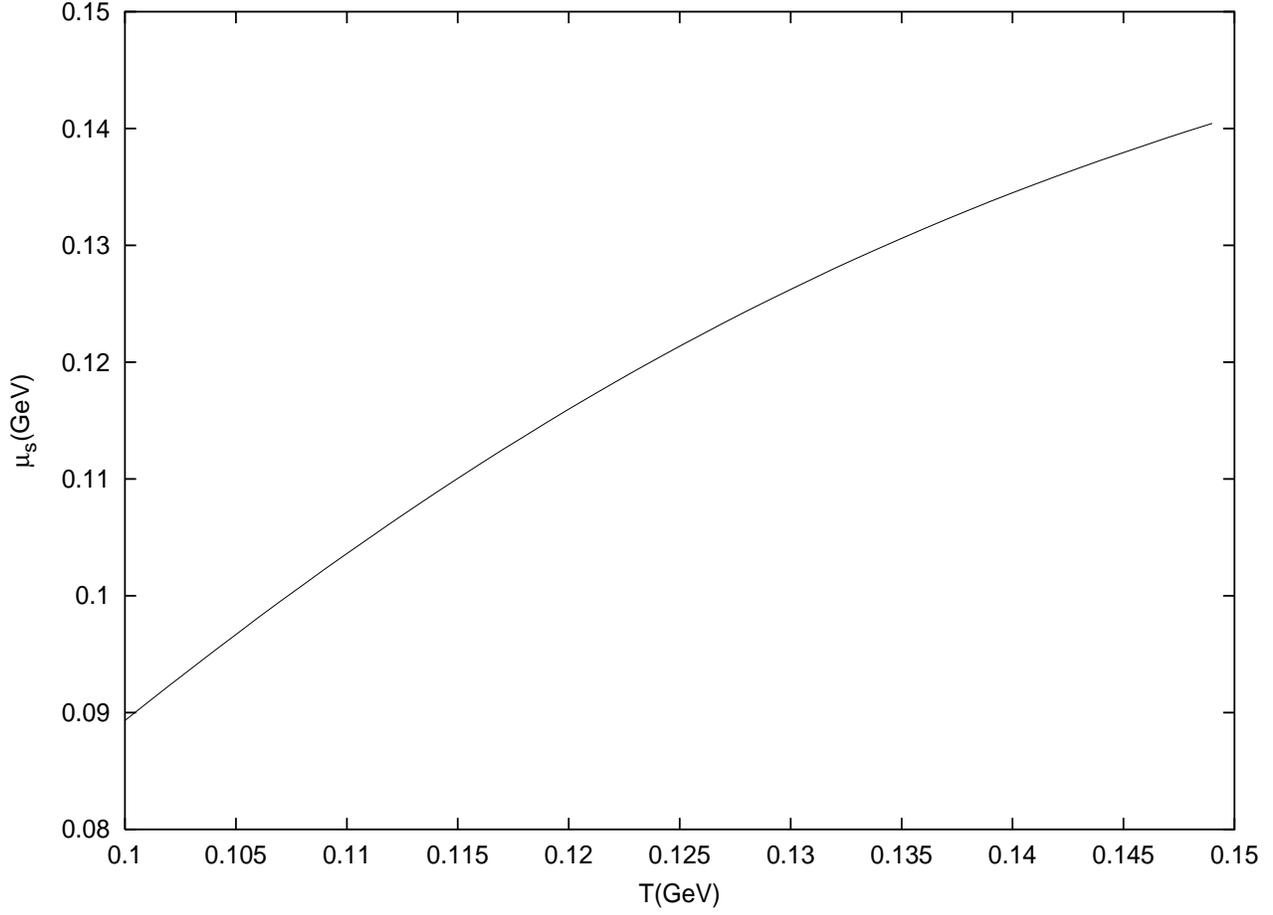}}}
\vskip 0.5cm \caption{ The dependence of $\mu_s$ on the
temperature T, for $\mu_B=0.55$ GeV and assuming $<s>$=0, for the
SIS-GSI experiment \cite{gsi-10gev}, Au+Au at 10 GeV/A. }
\label{figure7}
\end{figure}

\begin{figure}[tph]

\rotatebox{270}{\resizebox{356pt}{356pt}{\includegraphics[100,130][612,660]{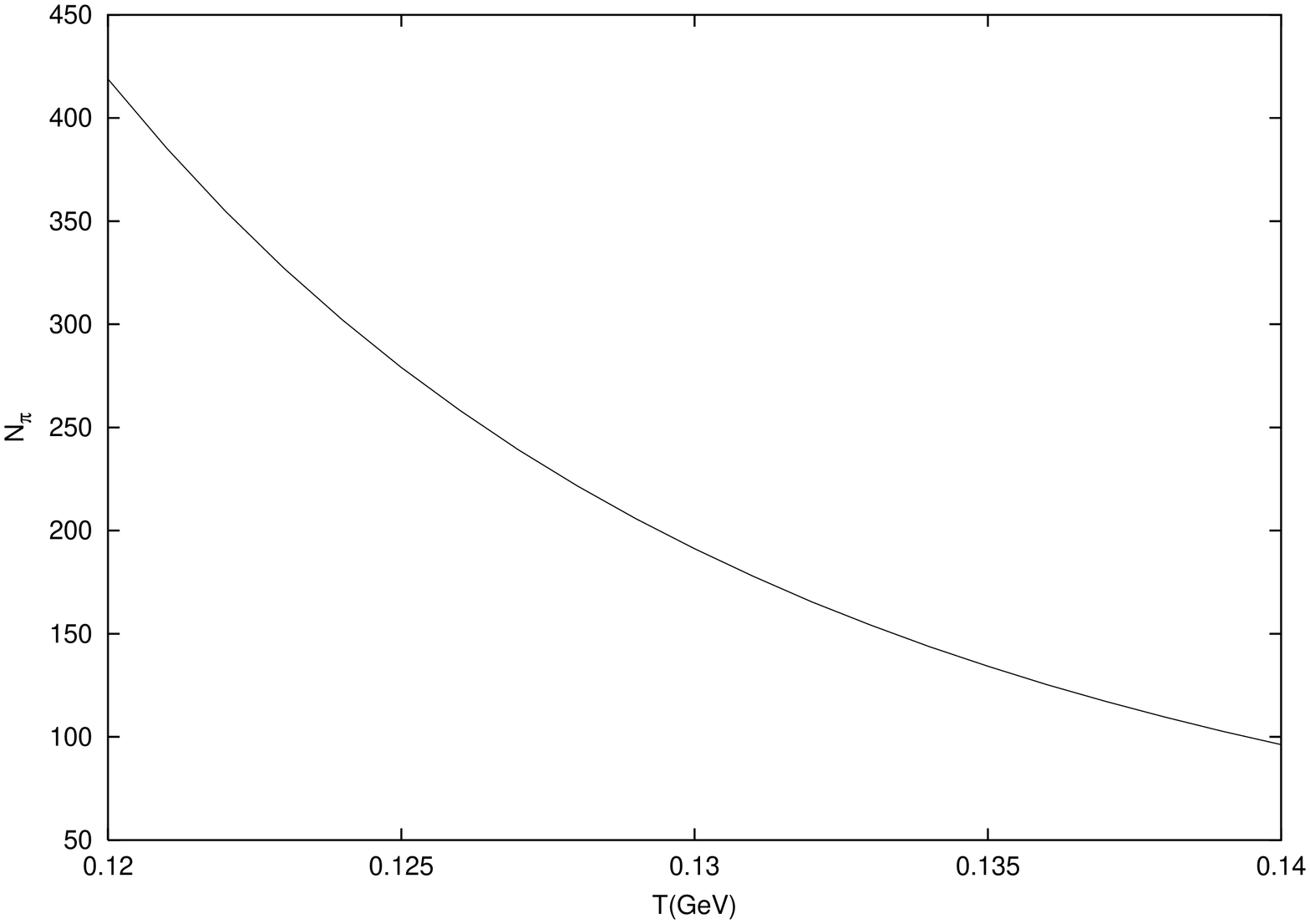}}}
\vskip 0.5cm \caption{ Total production rate of $\pi^+$ for Au+Au
at 10 GeV/A \cite{gsi-10gev}. For $T=0.13$ GeV the total $\pi^+$
production rate is approximately 180. The value $\mu_B=0.55$ GeV
was used in the calculations, as discussed in the text.}
\label{figure8}
\end{figure}

We have also determined ratios of particle production and some
absolute production rates. The particle production is calculated
for temperatures just below $T_c$, where only color zero states
are allowed. This implies that the partition function to use is
$Z_{(0,0)}$. We then apply Eq. (\ref{n-exp}). Note that in the
expression of the particle-production ratios the partition
function cancels out and only the dependence on the mass of the
particles and the chemical potential remains. Figure 9 shows
results for some particle-production ratios for beam energies
$\sqrt{s}=130$A GeV. The experimental values are taken from Ref.
\cite{rafelski1}, based on the experiment described in Ref.
\cite{exp-rafel} (see also \cite{braun-m}). For baryons, only the
ratios of particle and anti-particle production are shown because
these expressions are independent of the mass of the baryon. As
noted in (I) the masses of the baryons are not well reproduced
because they are considered as consisting of three idealized
fermions on top of the meson sea. The interaction to the meson sea
is not taken into account yet, but indications about how to do it
are given in (I).

\begin{figure}[tph]

\vskip 1cm
\rotatebox{270}{\resizebox{356pt}{356pt}{\includegraphics[100,130][612,660]{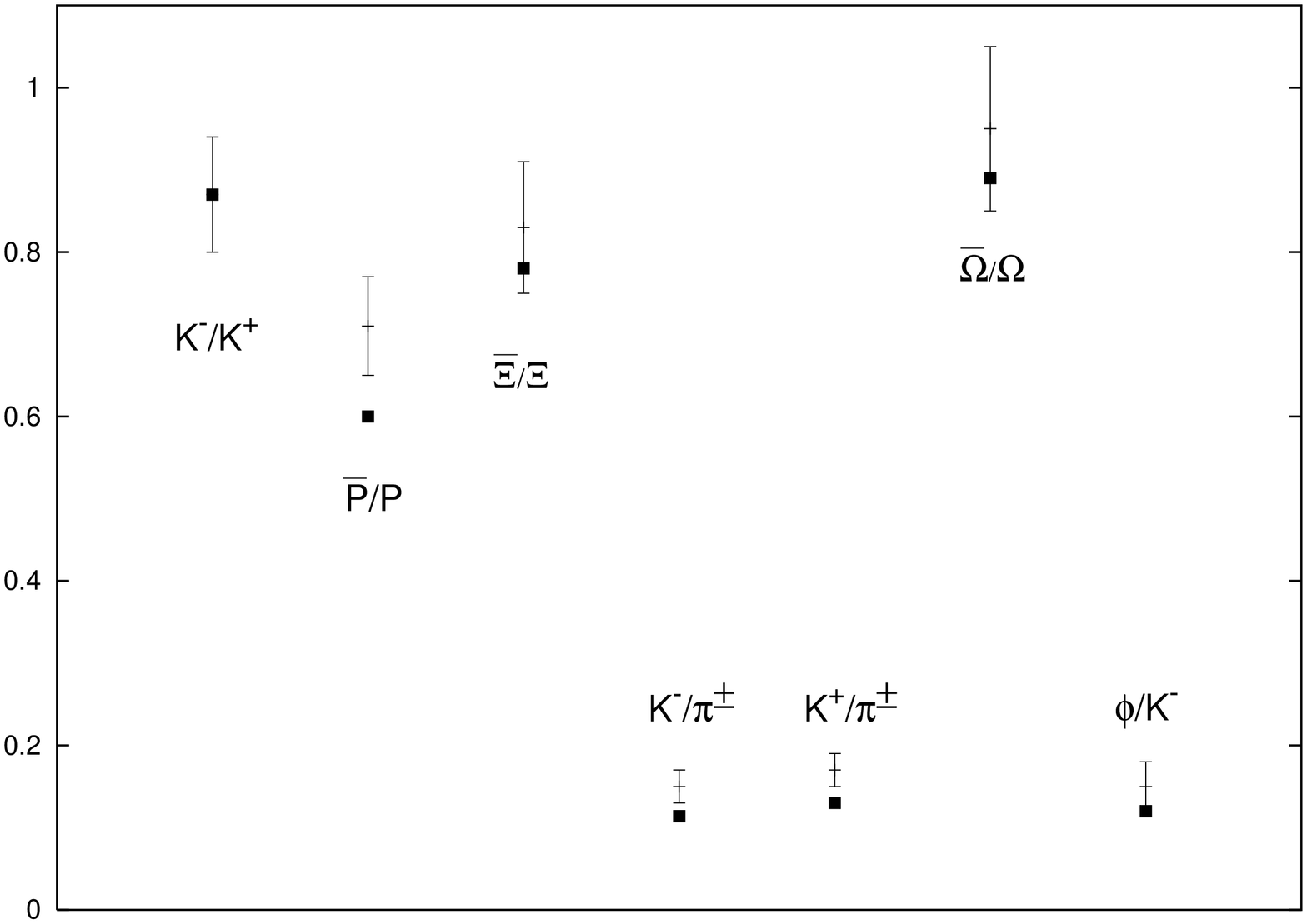}}}
%\vskip 0.5cm
\caption{ Some particle production ratios for the beam energies
 $\sqrt{s}=130$ A GeV taken from Ref. \cite{rafelski1}. The calculated values are shown
with full squares and the experimental values are shown within
error bars.} \label{figure9}
\end{figure}
The central value of the $K^-/K^+$ production ratio, shown in
Figure 9, was reproduced with the values $\mu_s=0.012$ GeV and
$\mu_B=0.044$ GeV. The other ratios are predicted by the model.
Considering the simplicity of the model, the ratios are found to
be in a reasonable agreement with data.

In order to obtain the total yields for kaons and for the $\pi^+$
pions it is necessary to introduce further assumptions about the
size of the QGP. The baryon density is given by
$\frac{<B>}{V_{el}}$, where $V_{el}$ is the size of the
representative volume, as explained before. In order to conserve,
on the average, the baryon number we multiply the baryon density
by the total volume and require that it must be equal to the total
baryon number, given by the number of participants $N_{part}$.
This leads to the total volume

\begin{equation}
V_{tot} = \frac{N_{part} V_{el}}{<B>_a} ~~~,
\label{vtot}
\end{equation}
where the index $a$ refers to, as in the partition function, color
($a=c$) when the average value is calculated in the QGP and
$a=(0,0)$ when it is calculated in the hadron gas. In the QGP the
average value $<B>_c$ before the transition is smaller than the
average value $<B>_{(0,0)}$ in the hadron gas after the
transition. This is due to the small value of $Z_{(0,0)}$ at
$T_c$, since many other possible color states are excluded from it
which do contribute to $Z_c$. As a consequence, for $T=T_c$ the
volume of the QGP phase, as a function of $N_{part}$, is much
smaller than the one in the hadron gas phase. Assuming a sphere,
the radius of the QGP phase is about 8 fm, and it changes to about
20 fm after the transition. This implies volumes of the order of
approximately 2 $ 10^3$ fm$^3$ and  3.4 $10^4$ fm$^3$,
respectively.

This transition is, as pointed out earlier, assumed to take place
suddenly at $T_c$ (probably it should be smeared out but we cannot
describe it with the present model). This implies that within the
scenario assumed there is a rapid expansion of the volume caused
by the transition from the QGP to the hadron phase, which should
be observed as a large outward motion. One possible interpretation
is that most of the pions are produced during the transition,
liberating energy and provoking a rapid expansion of the system.
The energy gained is represented by the jump between the lower and
upper curves of Figure 5, but its origin cannot be explained by
the present model, where confinement was shifted by hand.

Figure 10 shows the total pion yield as a function of the
temperature $T$, corresponding to the Au+Au collision. The upper
curve describes the total yield when all nucleons participate,
while for the lower one we have taken $N_{part}=250$. This value
agrees better with the experiment, as seen from the results, and
it means that in the collision about 250 nucleons participate in
the QGP. In Figure 11 the total kaon production rate is displayed.
The upper curve corresponds to $K^+$ and the lower one to $K^-$
absolute production rates, respectively. In both cases
$N_{part}=250$ was used, the same value used previously in the
calculation of the pion yield. The ratio of the curves was already
adjusted at the point corresponding to the $K^+/K^-$ ratio. The
absolute production rate and the shape of the curve, however, is a
prediction of the model (as far as we can talk about "predictions"
within this toy model). Considering the simplicity of the model it
is surprising that the absolute production rate is well
reproduced. This feature is common to other thermodynamical
descriptions of the transition from the QGP to the hadron-gas
\cite{braun-m,rafelski1}.

\begin{figure}[tph]

\vskip 2cm
\rotatebox{270}{\resizebox{356pt}{356pt}{\includegraphics[100,130][612,660]{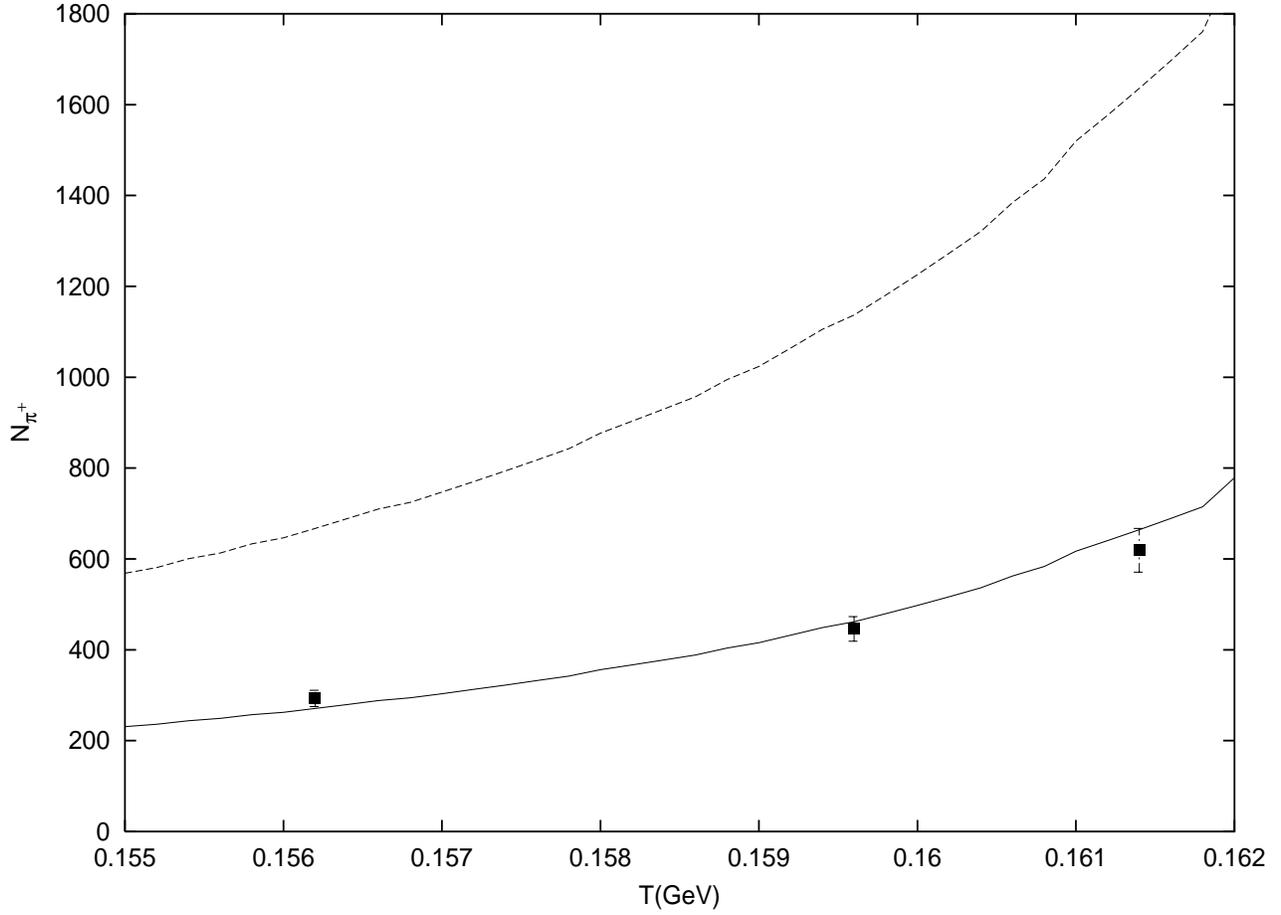}}}
\caption{ Total production rate of $\pi^+$. The upper curve is for
$N_{part}=394$ (Au+Au) and the lower one is for $N_{part}=250$.
Data are taken from \cite{piony}. Because of $\mu_I=0$, the
production rate for $\pi^-$ coincides with the production rate for
$\pi^+$. } \label{figure10}
\end{figure}

\begin{figure}[tph]

\rotatebox{270}{\resizebox{356pt}{356pt}{\includegraphics[100,130][612,660]{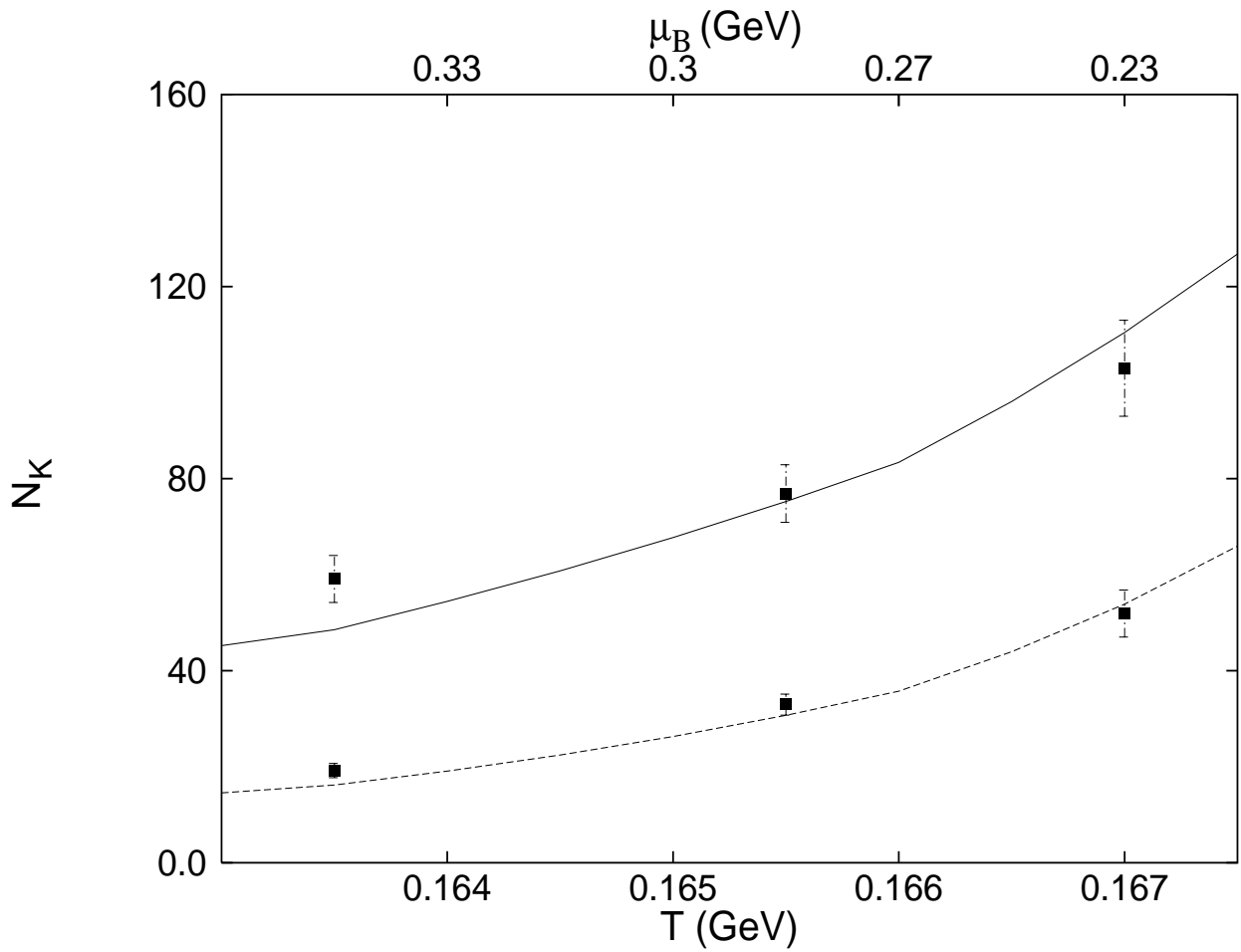}}}
\vskip 0.5cm \caption{ Total production rate of $K^+$ (upper
curve) and $K^-$ (lower curve) for $N_{part}=250$. Data are taken
from \cite{piony}. } \label{figure11}
\end{figure}

Finally, in Figure 12, we show the calculated expectation values
of the number of quark and gluon pairs as a function of the
temperature T. At $T=0$ GeV the results correspond to the
fractions of gluon pairs and fermion (quark-antiquark) pairs in
the physical vacuum state. At high temperatures the gluon part
increases and takes over the fermion part, which shows saturation.
However, at the temperatures of interest, i.e. around the point of
the phase transition $T_c \approx 0.16 $ GeV, the gluon number is
still suppressed with respect to the fermion pair number. This
might be in favor of the ALCOR model \cite{alcor} which supposes a
suppression of gluons in the QGP and takes only constituent quarks
and antiquarks into account. Note, that at $T=0.170$ GeV still a
sensible amount of gluon pairs are present.

\begin{figure}[tph]

\rotatebox{0}{\resizebox{656pt}{656pt}{\includegraphics[200,130][612,660]{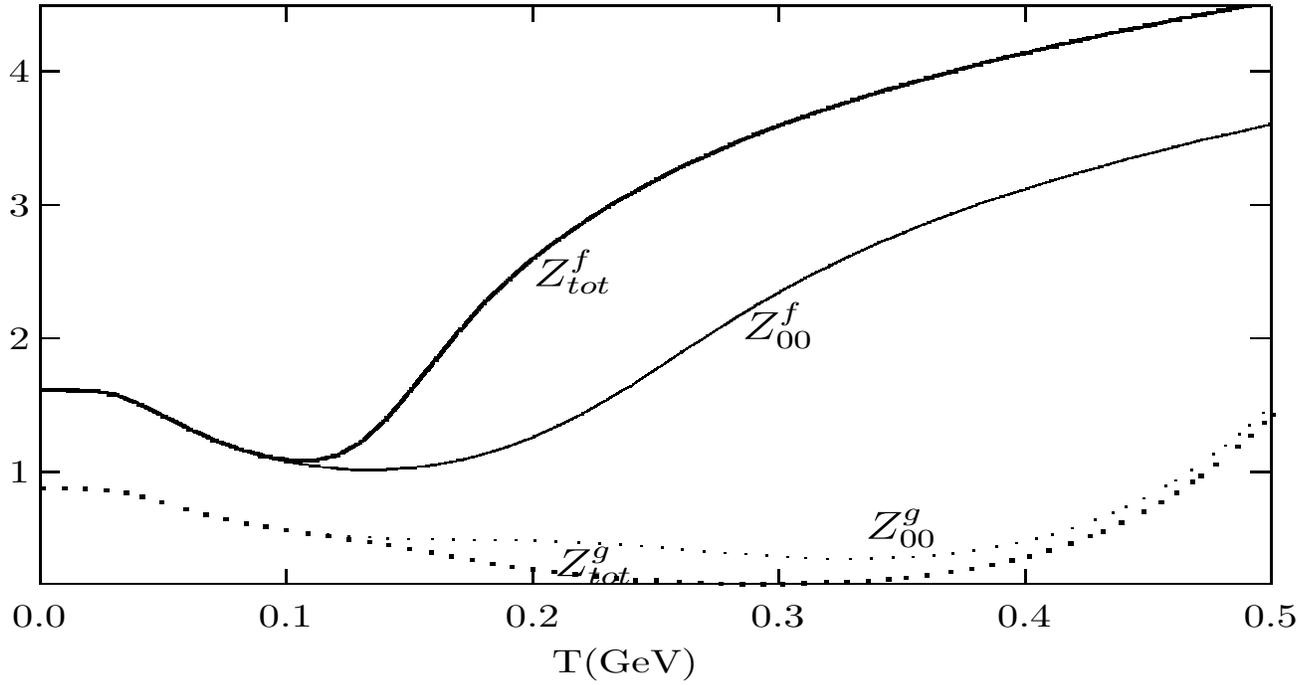}}}
\vspace{-10cm} \caption{ Expectation value of the fermion
(quark-antiquark) and gluon pairs. The symbols $Z_{tot}$ and
$Z_{00}$, on each curve, indicate if color was allowed. The upper
indices $f$ and $g$ on the $Z$ refer to fermion and gluon pairs
respectively. } \label{figure12}
\end{figure}

\section{Conclusions}

We have presented a toy model of QCD. The model is described in
(I) and in this paper we have focused on the thermodynamic
properties, at equilibrium, emerging from the model. We have
calculated the partition function with and without color, and
studied the temperature dependence of some observables, like the
internal energy, the heat capacity, and the production rates of
particles. The parameters of the model were determined in (I),
adjusting the meson spectrum at low energy. Without further
parameters the internal energy, the heat capacity and some
particle ratios were determined, as explained in the text.

We have applied the model to the case of the Au+Au collision at 10
GeV/A \cite{gsi-10gev} and shown that it can reproduce
qualitatively the absolute production rate of $\pi^+$. At this
energy the QGP has not yet formed and, therefore, the results show
that the model can be applied to study schematically the
thermodynamics of a hadron gas.

Next, we have applied the model to energies where one assumes that
the QGP has been already formed. The absolute production rate of
$\pi^+$ and kaons were calculated, just below the transition
temperature, by taking the number of participant nucleons
($N_{part}$) as an input. The agreement between calculated and
experimental values was found to be satisfactory. Also, the
resulting production rate was described reasonable well, once the
$\mu_s$ chemical potential was fixed to yield the correct
(observed central value) $\frac{K^+}{K^-}$ ratio. Some
mass-independent baryon-antibaryon ratios, were qualitatively
reproduced by the model predictions.

This demonstrates that the model is able to describe the general
trend of QCD, in the finite temperature domain, and the transition
to and from the quark gluon plasma.

\section{Acknowledgment}
We acknowledge financial support through the CONACyT-CONICET
agreement under the project name {\it Algebraic Methods in Nuclear
and Subnuclear Physics} and from CONACyT project number 32729-E.
(S.J.) acknowledges financial support from the {\it Deutscher
Akademischer Austauschdienst} (DAAD) and SRE, (S.L) acknowledges
financial support from DGEP-UNAM. Financial help from DGAPA,
project number IN119002, is also acknowledged.

\end{document}